\documentclass[a4paper,fleqn,usenatbib]{mnras}

\usepackage[T1]{fontenc}
\usepackage{ae,aecompl}

\usepackage{graphicx}	
\usepackage{amsmath}	
\usepackage{amssymb}	

\title[S2CLS: Deep number counts and the $z$-dist of the recovered CIB]{The SCUBA-2 Cosmology Legacy Survey: The EGS deep field I-- 
Deep number counts and the redshift distribution of the recovered Cosmic Infrared Background at 450 and 850\! $\bmath{\mu}$m}

\author[J. A. Zavala et al.]{
J. A. Zavala,$^{1}$\thanks{E-mail: \href{mailto:zavala@inaoep.mx}{zavala@inaoep.mx}}
I. Aretxaga,$^{1}$
J. E. Geach,$^{2}$
D. H. Hughes,$^{1}$
M.  Birkinshaw,$^{3}$
\newauthor
E. Chapin, $^{4}$
S. Chapman,$^{5}$
Chian-Chou Chen,$^{6}$
D. L. Clements,$^{7}$
J. S. Dunlop,$^{8}$
\newauthor
D. Farrah, $^{9}$
R. J. Ivison,$^{8,10}$
T. Jenness,$^{11,12}$
M. J. Micha{\l}owski,$^{8}$
E. I. Robson$^{8,13}$
\newauthor
Douglas Scott,$^{14}$
J. Simpson,$^{8}$
M. Spaans,$^{15}$
P. van der Werf$^{16}$
\\
$^{1}$Instituto Nacional de Astrof\'{i}sica, \'{O}ptica y Electr\'{o}nica (INAOE), Luis Enrique Erro 1, Sta. Ma. Tonantzintla, Puebla, Mexico\\
$^{2}$Center for Astrophysics Research, Science \& Technology Research Institute, University of Hertfordshire, Hatfield, AL10 9AB, UK\\
$^{3}$H. H. Wills Physics Laboratory, University of Bristol, Tyndall Avenue, Bristol BS8 1TL\\
$^{4}$Herzberg Astronomy and Astrophysics, National Research Council Canada, 5071 West Saanich Road, Victoria, BC V9E 2E7, Canada\\
$^{5}$Department of Physics and Atmospheric Science, Dalhousie University, 6310 Coburg Rd., Halifax, NS B3H 4R2, Canada\\
$^{6}$Centre for Extragalactic Astronomy, Department of Physics, Durham University, South Road, Durham DH1 3LE, UK\\
$^{7}$Astrophysics Group, Imperial College London, Blackett Laboratory, Prince Consort Road, London SW7 2AZ, UK\\
$^{8}$Institute for Astronomy, University of Edinburgh, Royal Observatory, Blackford Hill, Edinburgh EH9 3HJ, UK\\
$^{9}$Department of Physics, Virginia Tech, Blacksburg, VA 24061, USA\\
$^{10}$European Southern Observatory, Karl Schwarzschild Strasse 2, Garching, Germany\\
$^{11}$Joint Astronomy Centre, 660 N.A`oh\={o}k\={u} Place, University Park, Hilo, HI 96720, USA\\
$^{12}$Large Synoptic Survey Telescope Project Office, 933 N. Cherry Ave, Tucson, AZ 85721, USA\\
$^{13}$UK Astronomy Technology Centre, Royal Observatory, Blackford Hill, Edinburgh EH9 3HJ\\
$^{14}$Department of Physics \& Astronomy, University of British Columbia, 6224 Agricultural Road, Vancouver, BC, V6T 1Z1, Canada\\
$^{15}$Kapteyn Astronomical Institute, University of Groningen, Postbus 800, NL-9700 AV Groningen, the Netherlands\\
$^{16}$Leiden Observatory, Leiden University, PO Box 9513, 2300 RA Leiden, the Netherlands
}

\date{Accepted 2016 October 11. Received 2016 October 7; in original form 2016 June 2.}

\pubyear{2016}

\begin{document}
\label{firstpage}
\pagerange{\pageref{firstpage}--\pageref{lastpage}}
\maketitle

\begin{abstract}
We present deep observations at 450 $\mu\rm m$ and 850  $\mu\rm m$ in the  Extended Groth Strip field taken with the SCUBA-2 camera mounted on the 
James Clerk Maxwell Telescope as part of the deep
SCUBA-2 Cosmology Legacy Survey (S2CLS), achieving a central instrumental depth of $\sigma_{450}=1.2$\! mJy\! beam$^{-1}$ and $\sigma_{850}=0.2$\! mJy\! beam$^{-1}$. 
We detect 57 sources at 450\! $\mu\rm m$ and 90 at 850\! $\mu\rm m$ with S/N\! $>3.5$ over $\sim 70$ arcmin$^2$. From these detections we derive
the number counts at flux densities $S_{450}>4.0$\! mJy and $S_{850}>0.9$\! mJy, which represent the deepest number counts
at these wavelengths derived using directly extracted sources from only blank-field observations with a single-dish telescope. Our measurements smoothly
connect the gap between previous shallower blank-field single-dish observations and deep interferometric ALMA results. 
We estimate the contribution of our SCUBA-2 detected galaxies to the cosmic infrared background (CIB), as well as the contribution of 
24\! $\mu\rm m$-selected galaxies through a stacking technique, which add a total of $0.26\pm0.03$ and $0.07\pm0.01$\! MJy\! sr$^{-1}$, 
at 450\! $\mu\rm m$ and 850\! $\mu\rm m$, respectively.
These surface brightnesses correspond to $60\pm20$ and $50\pm20$ per cent of the total CIB measurements, where the errors are dominated by those of the total CIB. 
Using the photometric redshifts of the 24\! $\mu\rm m$-selected sample and the redshift distributions of the submillimetre galaxies, 
we find that the redshift distribution of the recovered CIB is different at each wavelength, with a peak at $z\sim1$ for 450\! $\mu\rm m$
and at $z\sim2$ for 850\! $\mu\rm m$, consistent with previous observations and theoretical models.
\end{abstract}




\section{Introduction}

Early studies of the cosmic infrared background (CIB) showed that the Universe emits a comparable energy
density at infrared (IR) and submillimetre wavelengths as it does at optical and ultraviolet wavebands, which
suggests that roughly half of the star light emission is absorbed and re-emitted  by dust in galaxies 
(e.g. \citealt{1987ARA&A..25..187S}; \citealt{1996A&A...308L...5P}; \citealt{1998ApJ...508..123F}). 
The first attempt to resolve this background resulted in the discovery of a new population of high-redshift ($z\sim 2-3$) 
galaxies (hereafter submillimetre galaxies, SMGs) through single-dish telescope observations at submillimetre wavelengths 
(e.g.  \citealt{1997ApJ...490L...5S}; \citealt{1998Natur.394..248B}; \citealt{1998Natur.394..241H}). 

The discovery of copious numbers of SMGs has 
proven to be a significant challenge for theoretical models of galaxy evolution. However, despite their large 
far-infrared (FIR) luminosities ($\ga10^{12}$ L$_\odot$) and their large space density (higher than local ULIRGs), these galaxies only represent a 
fraction of the measured CIB (see reviews by \citealt{2002PhR...369..111B} and \citealt{2014PhR...541...45C}). 
Previous blank field surveys resolved 20--40 per cent of the CIB at 850\! $\mu\rm{m}$ (e.g., \citealt{2000AJ....120.2244E}; \citealt{2006MNRAS.372.1621C}).
Meanwhile, in cluster fields 50--100 per cent of the CIB is resolved, thanks to the effect of gravitational lensing 
(e.g. \citealt{1997ApJ...490L...5S}; \citealt{2002AJ....123.2197C}; \citealt{2008MNRAS.384.1611K}; \citealt{2013ApJ...776..131C}),
suggesting that a large fraction of this background originates from faint
sources ($S_{850}< 3$\! mJy) whose number counts are not yet well constrained. At shorter wavelengths, closer to the peak of the CIB 
($\lambda \approx200$\! $\mu\rm{m}$;  \citealt{1998ApJ...508..123F}), observations with the {\it Herschel Space Observatory} have resolved $\sim75$ per cent
of the CIB at 100 and 160\! $\mu\rm{m}$ (\citealt{2011A&A...532A..49B}; \citealt{2013A&A...553A.132M}). Nevertheless, at 250, 350 and 500\! $\mu\rm{m}$ only 
a small fraction ($<25$ per cent; \citealt{2010A&A...518L..21O, 2012MNRAS.424.1614O}) has been resolved in individual galaxies, 
due to its much higher confusion limit, although, using statistical methods such as stacking and P(D), a significant larger fraction 
($\approx40-70$ per cent) can be associated with discrete sources (e.g. \citealt{2010MNRAS.409..109G}; \citealt{2012A&A...542A..58B}; \citealt{2013ApJ...779...32V}), 
with recent study suggesting a resolved fraction greater than 90 per cent (\citealt{2015ApJ...809L..22V}; see also \S\ref{sec:CIB}).

Consequently, deeper observations with higher angular resolution  are still required to directly resolve and to completely understand 
the nature of this emission, as well as to fully determine the number counts of SMGs at fainter flux densities. These are some of 
the key science drivers for the SCUBA-2 Cosmology Legacy Survey (S2CLS), which exploits the  capabilities of the SCUBA-2 camera \citep{2013MNRAS.430.2513H}, 
efficiently achieving large and deep (confusion limited) maps at both 450 and 850\! $\mu\rm{m}$, simultaneously. At the shorter wavelength, the angular resolution is 
$\theta_{\rm FWHM}\sim 8$\! arcsec, a factor of about 5 better than {\it Herschel} at $500$\! $\mu\rm{m}$ ($\theta_{\rm FWHM}\sim 36.6$\! arcsec), 
which results in a confusion limit around 7 times deeper.

On the other hand, the interferometric technique combined with the high sensitivity of ALMA has allowed the exploration of a range of flux densities unreachable 
through single-dish telescopes (e.g. \citealt{2013ApJ...769L..27H}; \citealt{2014ApJ...795....5O}; \citealt{2015A&A...584A..78C}; \citealt{2015arXiv150805099O};
\citealt{2016arXiv160706769A}; \citealt{2016arXiv160600227D}; \citealt{2016ApJS..222....1F}). However, because of its relatively small field of view, it is 
observationally expensive to map large areas of blank sky with ALMA, and therefore difficult to constrain the less abundant population of bright galaxies. For these reasons, 
deep single dish telescope observations are still necessary to bridge the gap between new deep interferometric results and the past single-dish shallower studies. 

Here we present 450 and 850\! $\mu\rm{m}$ observations taken in the Extended Groth Strip (EGS) field as part of the deep tier of the S2CLS 
(the wide tier of the survey has been presented by \citealt{2016arXiv160703904G}). This field has been the target of the 
All-wavelength Extended Groth strip International Survey (AEGIS), which includes observations of some of the world's most  powerful telescopes, from X-rays to radio wavelengths.
James Clerk Maxwell Telescope (JCMT) observations were scheduled to take advantage of excellent conditions ($\tau_{\rm 225 GHz} <0.05$) at the top of Manua Kea, which results in the deepest single-dish
telescope blank-field observations achieved at these wavelengths, comparable to the deep S2CLS maps in COSMOS and UDS
(\citealt{2013MNRAS.432...53G}; \citealt{2013MNRAS.436..430R}; \citealt{2016MNRAS.458.4321K}).

In this paper we report the first results of the EGS deep study. We describe the observations and data reduction in \S2. The maps and source extraction
are described in \S3. In \S4 we report the estimated number counts at each wavelength, and in \S5 the  CIB fraction recovered. Finally, our
results are summarized in \S6. The multi-wavelength analysis, as well as a description of the physical properties of these galaxies will be presented
in a subsequent paper (Zavala et al. in preparation).

All calculations assume a standard $\Lambda$ cold dark matter cosmology with $\Omega_\Lambda=0.68$, $\Omega_{\rm m}=0.32$, and
$H_0=67$ kms$^{-1}$Mpc$^{-1}$ (\citealt{2014A&A...571A..16P}).

\section{Observations and data reduction}

Observations at 450 and 850\! $\mu\rm{m}$ were taken, simultaneously, with the SCUBA-2 camera on the JCMT
between 2012 and 2015 under the best weather conditions (band 1, $\tau_{\rm 225 GHz} < 0.05$) as part of the deep S2CLS in the  
extragalactic EGS field. A standard  $\sim$5 arcmin diameter  `DAISY' mapping pattern (\citealt{2014SPIE.9153E..03B}) was used, 
which keeps the pointing centre on one of the four SCUBA-2 sub-arrays during the scanning. All data were reduced using 
the standard {\sc SMURF}  package (\citealt{2011ASPC..442..281J}; \citealt{2013MNRAS.430.2545C}) with the default `blank field'
configuration, although the procedure for map-making (standardized across the S2CLS project) differs slightly from that described
in Section 4.2 of \citet{2013MNRAS.430.2545C}. This process is described in detail in \citet{2013MNRAS.432...53G} and 
\citet{2016arXiv160703904G} which we summarize here, emphasizing the differences with \citet{2013MNRAS.430.2545C}. 

The signal recorded by each bolometer is assumed to be a linear combination of atmospheric emission, astronomical signal (attenuated 
by atmospheric extinction), and a noise term. While extinction may be corrected directly using external measurements (i.e., extrapolating 
from $\tau_{\rm 225 GHz}$ measured at the adjacent Caltech Submillimeter Observatory), the dynamic iterative map maker attempts to solve for the
remaining components, refining the model until convergence is met, at which either an acceptable tolerance has been reached, or a fixed 
number of iterations has been completed. First, all of the bolometer data are re-sampled to a lower data rate 
corresponding to the Nyquist frequency for the chosen pixel size, and filtered
to retain only frequencies relevant to point-sources scales, which results in 
a band-pass filter defined as $v/150\ {\rm arcsec} < f < v/4\ {\rm arcsec}$, where $f$ is in Hz, 
and $v$ is the scan speed in $\rm arcsec\ s^{-1}$. Then, each iteration 
consists of the following essential steps: (i) removing inter-bolometer correlated noise (primarily sky emission) via common-mode suppression
(subtracting a time-varying template created from the average of all bolometer signals in a given subarray, using per-bolometer gains and 
offsets to fit it to the signal in question prior to removal); (ii) producing maps (on 2\! arcsec $\times$ 2\! arcsec pixel grids) from the resulting 
time streams which should, ideally, consist only of astronomical sources and higher-frequency (un-correlated) noise; and finally (iii) 
re-projecting the map back into the time-domain to estimate the contribution of astronomical sources to the bolometer signals (e.g., 
`scanning' the detectors across the current map estimate), and then subtracting these signals, leaving primarily high-frequency, un-correlated
noise from which a time-domain variance can be measured for each bolometer. The variance measured in step (iii) is used to weight the data 
in step (ii) in subsequent iterations. We also note that the common-mode subtraction step in (i) provides an efficient mechanism for flagging 
bad data: portions of bolometer signals that do not resemble the common-mode are simply masked and ignored in the remaining analysis. Since the
signal-to-noise ratio (S/N) of astronomical sources is low in these fields (generally undetectable in a single bolometer signal), the solutions converge quickly. 
The map-maker halts when the reduced $\chi^2$ changes by less than 0.05, or a maximum of 20 iterations has been reached. We note that the maps 
are quite insensitive to the values of these convergence criteria provided that a handful of iterations are completed.

The maps produced by this procedure continue to exhibit weak large-scale noise features (though on scales smaller than the 150\,arcsec cutoff 
in the initial band-pass filter) due to any non-white noise that may have gotten through the common-mode subtraction step. \citet{2013MNRAS.430.2545C} 
advocates using jackknife maps at this stage to empirically measure this noise (i.e., dividing the data into two halves, producing maps of each,
and then taking their difference to yield a map containing only noise without astronomical sources), and then constructing a `whitening filter' 
to flatten the map to assist with source detection. While in some sense optimal, this procedure will produce a different effective PSF for each 
observation, making comparison between this field and others in the S2CLS more complicated. For this reason, across the S2CLS, we have opted to use 
a slightly more conservative (and uniform) method of large-scale noise suppression which involves subtracting a low-pass filtered map 
(accomplished by smoothing the map with a 30\,arcsec FWHM Gaussian kernel). This method has been used by virtually all groups analyzing fields of
point sources observed with SCUBA-2 data to date. In other areas of astronomy, this procedure is known as a linear unsharp mask.

Finally, in order to detect sources, we apply a matched filter to the maps, using an effective PSF constructed from an estimate of the
diffraction-limited SCUBA-2 beams (Gaussians, with $\theta_{\rm FWHM}\approx 8$ and 14.5\,arcsec for the 450 and 850\! $\mu\rm{m}$ bands, respectively, 
\citealt{2013MNRAS.430.2534D}), 
filtered using the same 30\,arcsec FWHM background subtraction kernel (which introduces negative sidelobes). Note that this procedure is only optimal
in the case that point sources are isolated in fields of uncorrelated white noise. A more sophisticated `confusion-compensating' matched filter 
was proposed in \citet{2011MNRAS.411..505C} in which the effects of source blending are included explicitly in the estimate of the noise power spectrum, 
which in turn leads to a point-source detection kernel that behaves as the PSF described here in the low-S/N regime (when considering only instrumental noise),
and smoothly converges to 
a de-convolution operator in the limit of infinite S/N. However, since this more optimal filter is a function of both the S/N of the observations
and an estimate of the source counts in the field, it would again lead to complications when comparing different S2CLS fields. For this 
reason, we have opted for this simpler, uniform source-finding kernel across the project.

\subsection{Astrometry}

\begin{figure}
  \includegraphics[width=\columnwidth]{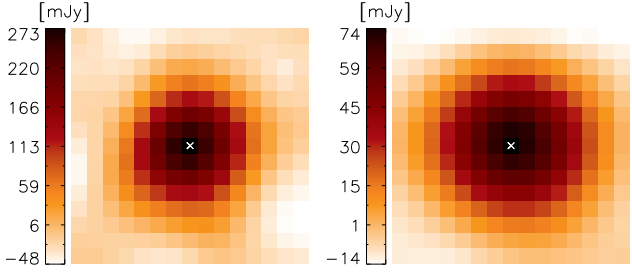}
  \caption{30\! arcsec $\times$ 30\! arcsec co-added postage stamps of the SCUBA-2  450 and 850\! $\mu\rm{m}$  flux maps (left and right, 
  respectively)  extracted from regions centred at 39 radio source positions that lie within the SCUBA-2 map boundary 
  \citep{2007ApJ...660L..77I}.  The stacked signal peaks at the central pixel for each map (white cross) which means that the astrometry of the
  observations is better than the 2\! arcsec pixel-size. The FHWM of the best-fit  Gaussian at 450\! $\mu\rm{m}$ is  11\! arcsec $\times$ 12\! arcsec,
  and 14.5\! arcsec $\times$ 16\! arcsec at 850\! $\mu\rm{m}$. 
  The $\sim$30 per cent broadening over the nominal 450\! $\mu\rm{m}$ FWHM of the PSF can be explained by a multiplicity of 20--30 per cent 
  (as previously reported by interferometric results, e.g. \citealt{2007MNRAS.380..199I}; \citealt{2011ApJ...726L..18W}; \citealt{2013ApJ...768...91H}) where the secondary 
  source only contributes $\sim20$ per cent of the total flux density.}
  \label{stackVLA}
\end{figure}

The identification of radio counterparts has been used to improve the astrometric accuracy of the submillimetre maps and 
to measure positional uncertainty of SMGs (e.g. \citealt{2002MNRAS.337....1I}; \citealt{2005ApJ...622..772C}). We use the 
VLA/EGS 20-cm survey \citep{2007ApJ...660L..77I} for this purpose. The position of the sources in this catalogue are adopted to stack 
the signal in our beam-convolved SCUBA-2 maps. Figure \ref{stackVLA}  shows 30\! arcsec $\times$ 30\! arcsec postage stamps extracted from regions 
centred at the radio positions and stacked together. The stacked signal peaks at the 
central pixel in the co-added postage stamps, indicating that there is no systematic offset between the SCUBA-2 data and the radio catalogue, or it if 
exists, is less than our pixel size (2\! arcsec $\times$ 2\! arcsec). Therefore, no positional correction was applied.

Since the stacking could be dominated by a few bright sources, we repeated the procedure but re-normalizing each image to a constant
peak brightness, finding consistent results.

\section{Maps and source catalogue}

\subsection{Maps}

\begin{figure*}
  \includegraphics[width=185mm]{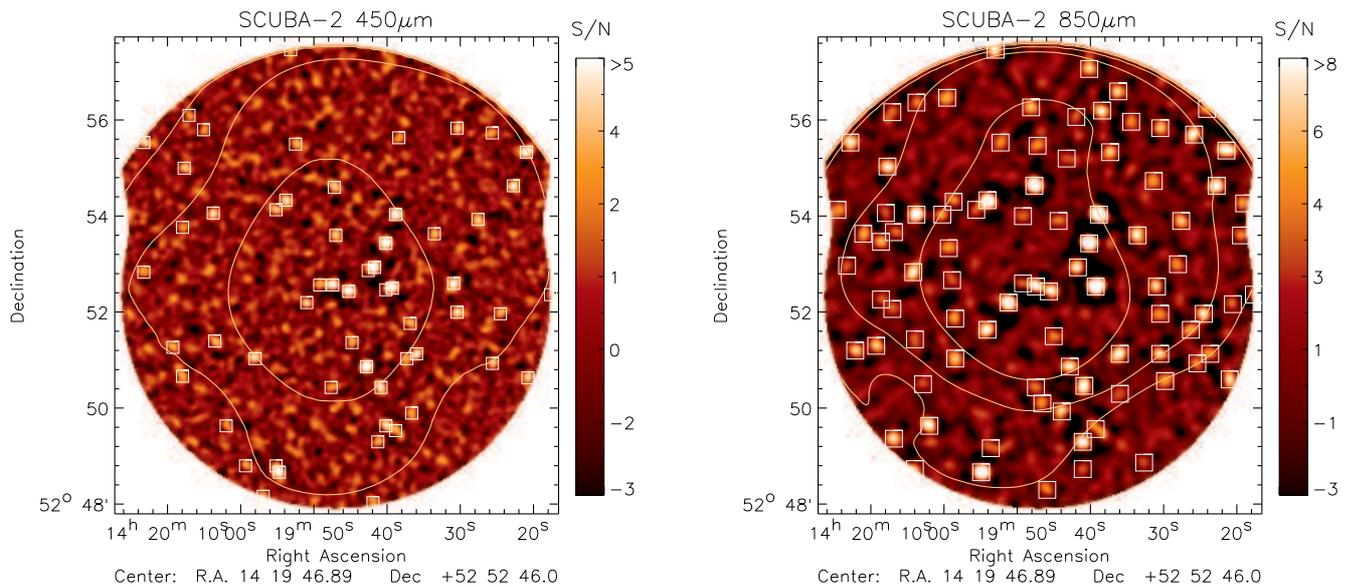}
  \caption{The 450 and 850\! $\mu\rm{m}$ Extended Groth Strip SCUBA-2 signal-to-noise ratio maps over an area of $\approx$70 arcmin$^2$. 
  The yellow contours show the variation in the noise level, and are spaced at 0.4 and 0.05\! mJy for the  450 and 850\! $\mu\rm{m}$ maps
  and starting at 1.5 and 0.25\! mJy, respectively. The identified S/N$>$3.5 sources candidates are marked with yellow squares 
  in both maps. The maps have been scaled to emphasize the visibility of the detected sources.}
  \label{mapaSN}
\end{figure*}

The 450 and 850\! $\mu\rm{m}$ signal-to-noise ratio  maps of EGS acquired with SCUBA-2 on the JCMT are shown in Figure \ref{mapaSN}. 
Each map  has a radially varying coverage (see contours in Figure \ref{mapaSN}), which is roughly uniform over the
central $\sim2$ arcmin  and increases radially towards the edge of the map as the effective exposure time decreases. The maximum instrumental depth 
achieved in the centre of each map is 1.2 and 0.2\! mJy\! beam$^{-1}$ at 450  and 850\! $\mu\rm{m}$, respectively. The noise has been estimated
through a jackknife procedure where 50 per cent of the individual scans are inverted. The total area considered for 
source extraction is $\approx 70$ arcmin$^2$, where the r.m.s noise is below 2.5 and 0.5\! mJy\! beam$^{-1}$, respectively.

\subsection{Source extraction and source catalogue}

To identify source candidates, we search for pixels in the (beam convolved) S/N map with values $>3.0$. If a peak is found, we 
adopt the celestial coordinates of the pixel, the flux density and the noise, and subtract the PSF estimated in the map-making procedure scaled 
to the S/N measured at this position. The process is repeated until there are no more pixels with S/N\! $>3.0$. 
All the source candidates above this threshold are  listed in Table \ref{catalogue}, together with their coordinates, 
measured S/N, raw flux densities, and  deboosted flux densities (see \S \ref{sec:deboosting}). However, a conservative threshold 
of S/N\! $>3.5$ has been adopted to define a more robust sample. The 3.5 threshold value is chosen to be the S/N 
level at which the contamination rate due to false detection is estimated to be less than 5 per cent at 850\! $\mu\rm{m}$ and less than 10 per cent at 
450\! $\mu\rm{m}$. At this threshold we detect 57 sources at 450 $\mu\rm m$ and 90 at 850 $\mu\rm m$. These sources are marked with squares 
in Figure \ref{mapaSN}.

\subsection{Completeness and positional uncertainty}\label{sec_completeness}
The detection rate for a given source within a flux density range is affected by both confusion noise from the underlying 
population of faint sources and instrumental Gaussian noise in the map. To account for these effects, we estimate 
the completeness of source detection using simulations in 
which mock sources with different flux densities are inserted into the observed EGS/SCUBA-2 signal maps, 
and then recovered with the same source extraction procedure used for the real catalogue. As 
described in detail by \citet{2008MNRAS.385.2225S, 2010MNRAS.405.2260S}, this method has the benefit of taking into account the effects of random and confusion noise
in the signal map and, since the sources are inserted one at a time, it does not modify the properties of the real map. 
We insert 10,000 sources in each flux density bin, ranging from 1 to 20\! mJy (in bins of 0.5\! mJy) for the 450\! $\mu\rm{m}$ map
and from 0.1 to 5\! mJy (in bins of 0.1\! mJy) for the 850\! $\mu\rm{m}$ map.
A source is considered recovered if it is extracted
with S/N\! $>3.5$ at a radius $<1.5$ times the size of the corresponding beam of its input position. Figure \ref{completeness} shows the completeness 
fraction as a function of flux density for this survey.

On the other hand, the relatively large beam--size combined with the low S/N of the detections results in a significant error on the position of the 
source candidates.  We characterize the positional uncertainty as a function of S/N from the same simulations,
where now we focus on the distance at which the artificial sources were extracted. 
The positions of the simulated sources are chosen randomly within a pixel (instead of adopting the pixel centre)
in order to take into account the 2\! arcsec $\times$ 2\! arcsec pixel size.
Figure \ref{completeness} also shows the median positional offset between the measured output source position and its input position as a function of S/N.

\subsection{Flux deboosting} \label{sec:deboosting}
When sources are detected at low S/N, their measured flux densities are systematically larger than their intrinsic flux densities if the number of 
sources increases while decreasing flux (e.g. \citealt{1998PASP..110..727H}). In this case it becomes more likely that the numerous faint sources are boosted high 
than the rarer bright sources to lower fluxes. This is particularly important in surveys of SMGs since the intrinsic population is known to have a 
steep distribution of counts, and also because the galaxies are typically detected at low S/N.

\begin{figure}
  \includegraphics[width=42mm]{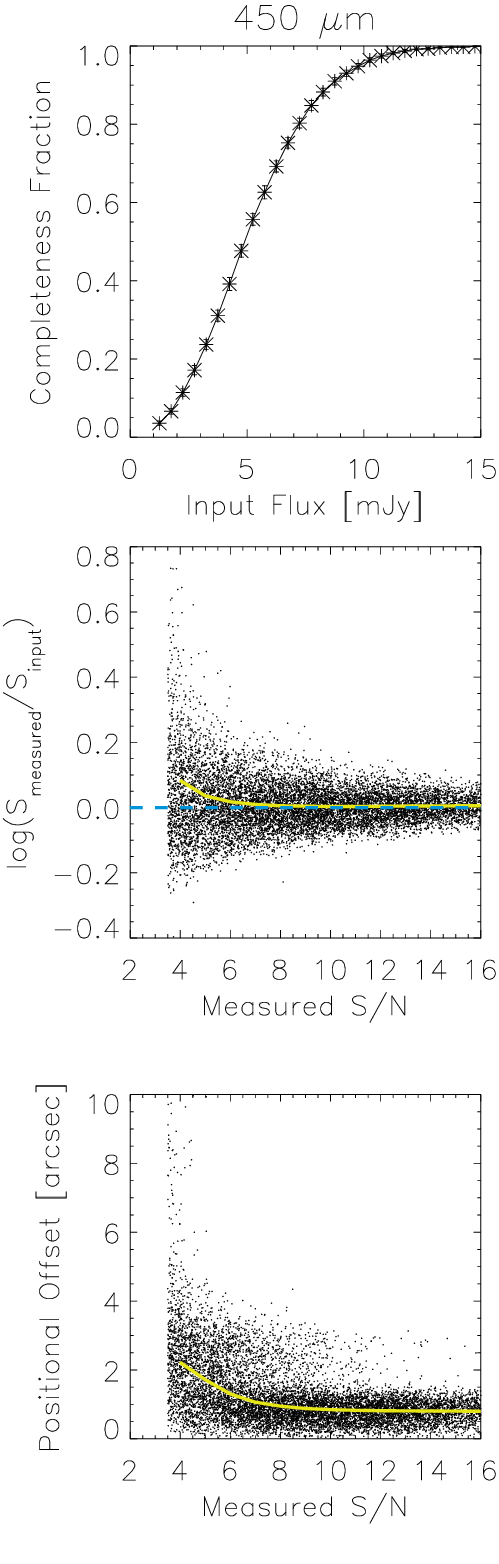}\includegraphics[width=42mm]{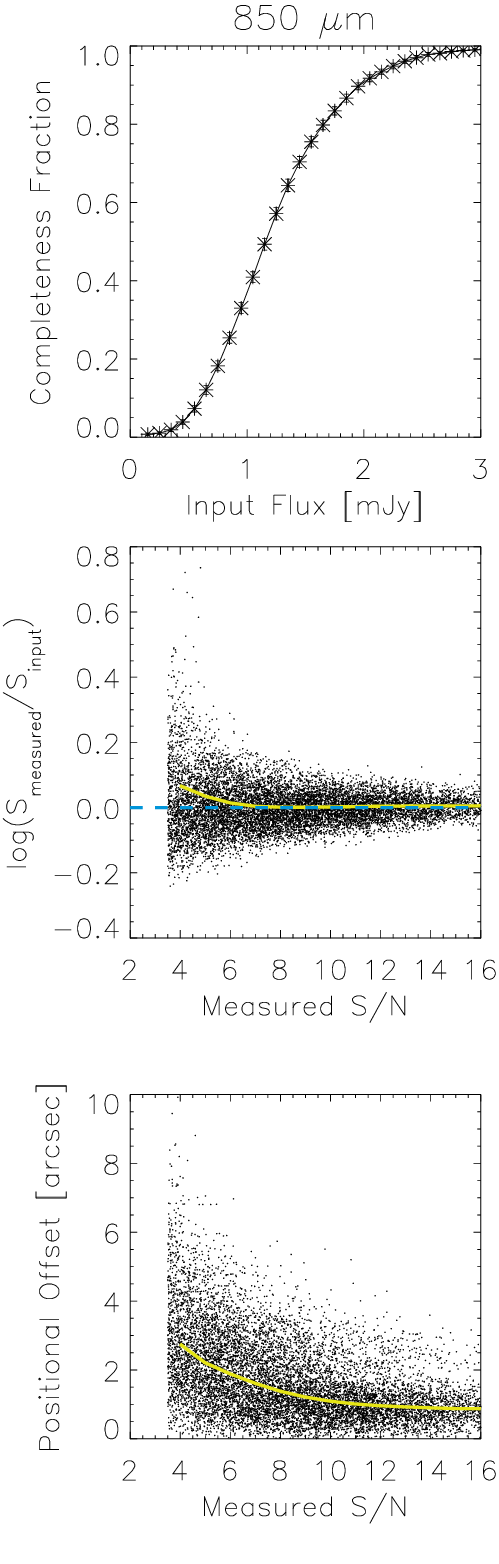}
  \caption{{\it Top}: Completeness of the 450 (left) and 850\! $\mu\rm{m}$ (right) source catalogues as a function of flux density. 
  The points and error bars show the completeness estimated by 
  inserting  sources of known flux density one at a time into the observed signal maps and then finding them with the same source extraction algorithm
  used to create the source catalogue. {\it Middle:} The boosting factor estimated from the same set of simulations, measured as the ratio of output measured 
  flux density and input flux density. The yellow solid line represents the median boosting factor in each bin of S/N and the blue dashed line represents 
  no flux boosting. {\it Bottom}: Positional uncertainty as a function of S/N, calculated as the offset between the measured output position and its input 
  position. The yellow line represents the median value in each bin of S/N.}
  \label{completeness}
\end{figure}

In addition to the completeness and positional uncertainties, the simulations described above allow us to calculate the boosting factor.
This `flux boosting' is measured as the ratio of the output measured flux density and the input flux density, as shown in Figure~\ref{completeness}.
The boosting factor can also be estimated as a function of both S/N and the local noise in the map, selecting only the simulated sources that 
have been inserted in specific regions within a specific noise range. This selection is important, as we know that our maps have radially varying 
sensitivity. The error in the deboosting factor is estimated as the standard deviation in each bin of S/N and noise, and then is propagated
to estimate the error in the deboosted flux density.

An alternative statistical method  has been developed to correct the flux boosting. For each source candidate we can estimate a posterior flux distribution
(PFD) which describes the intrinsic flux density of the source in terms of probabilities. The PFD is calculated using the Bayesian approach of 
\citet{2005MNRAS.357.1022C, 2006MNRAS.372.1621C}. For an individual source
detected with measured flux density $S_{\rm m} \pm \sigma_{\rm m}$, the probability distribution for its intrinsic flux density $S_{\rm i}$ is given by

\begin{equation}
  p(S_{\rm i}|S_{\rm m},\sigma_{\rm m})=\frac{p(S_{\rm i})p(S_{\rm m},\sigma_{\rm m}|S_{\rm i})}{p(S_{\rm m},\sigma_{\rm m})},
\end{equation}
where $p(S_{\rm i})$ is the prior distribution of flux densities, $p(S_{\rm m},\sigma_{\rm m}|S_{\rm i})$ is the likelihood of observing the data and 
$p(S_{\rm m},\sigma_{\rm m})$ is a normalization factor. We assume a Gaussian noise distribution for the likelihood of observing the data, where

\begin{equation}
  p(S_{\rm m},\sigma_{\rm m}|S_{\rm i})=\frac{1}{\sqrt{2\pi\sigma_{\rm m}^2}}\exp[\frac{-(S_{\rm m}-S_{\rm i})^2}{2\sigma_{\rm m}^2}]
\end{equation}
This assumption is justified by the Gaussian flux distribution observed in jackknife noise maps of our EGS data (see also \citealt{2013MNRAS.432...53G}).

The prior distribution of flux densities $p(S_{\rm i})$ is estimated by generating 10000 noiseless sky realizations, where sources are inserted with a uniform
spatial distribution into a blank map according to a number counts distribution.  Each source is  assume to be the PSF scaled by the flux density. The
pixel histogram of flux values from all these realizations gives an estimate of $p(S_{\rm i})$. We assume the prior number counts to be a Schechter function with the 
best-fitting parameters for the SCUBA-2 450 and 850\! $\mu\rm{m}$ COSMOS number counts from \citet{2013MNRAS.436.1919C} or \citet{2013MNRAS.432...53G}.
We find no significant differences between the PFDs of sources estimated when using these different priors.

\begin{figure}
  \includegraphics[width=75mm]{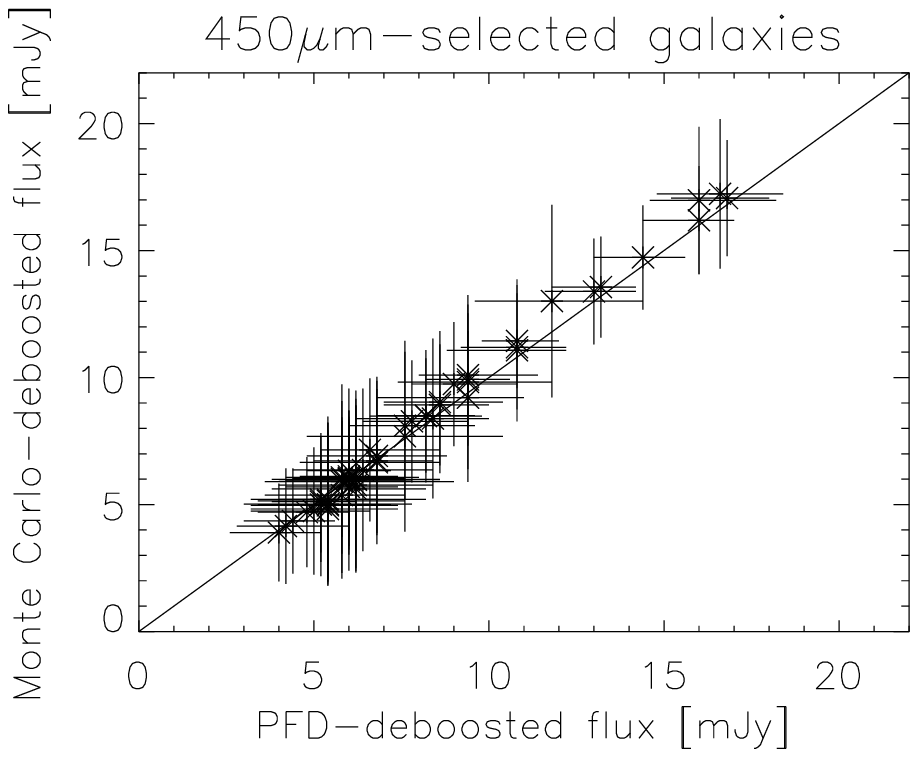}
  \includegraphics[width=75mm]{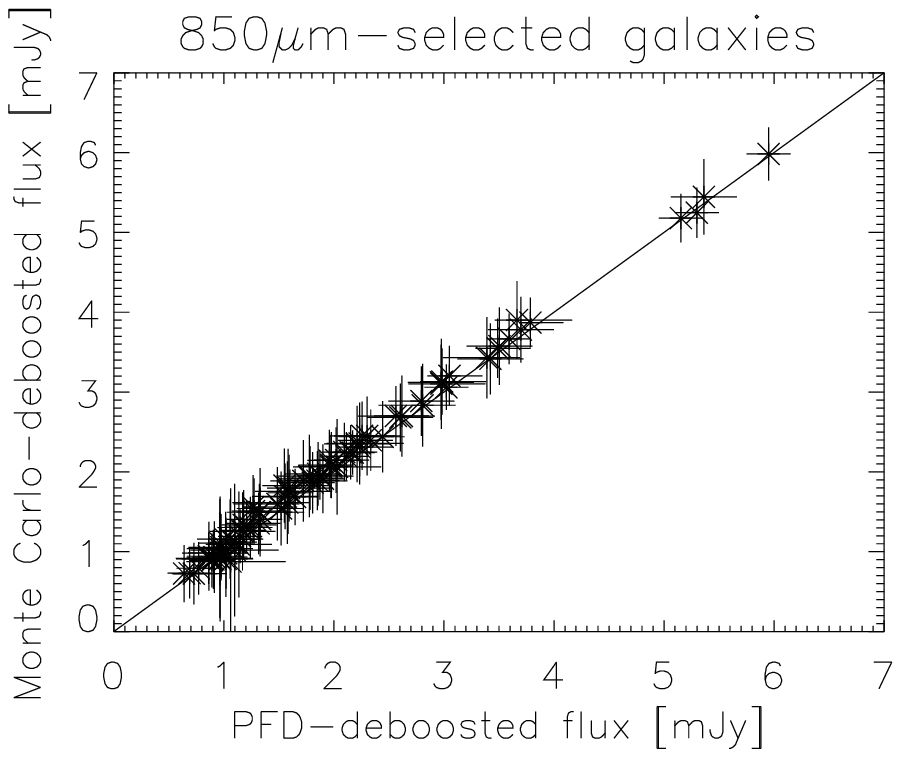}
  \caption{Comparison between the deboosted flux density of the source candidates detected at S/N$>3.5$ ({\it top: 450\! $\mu\rm{m}$}, 
  {\it bottom: 850\! $\mu\rm{m}$}), estimated through Monte Carlo simulations and the Bayesian method.}
  \label{comp_deboost}
\end{figure}

The deboosted flux density for each source is assumed to be the maximum value of the PFD and its associated 68
 per cent confidence interval. We have compared the deboosted flux density for each source estimated using both methods, i.e. Monte Carlo simulations and the Bayesian PFD. 
As we can see in Figure~\ref{comp_deboost}, the results of both methods are in good agreement. 

Finally, from the PFD, we measure the probability that each  source candidate will be deboosted to less than 0\! mJy.
This value could be used to exclude candidates that exceed some probability threshold, and therefore decrease the contamination 
of false sources in our catalogue (see Section \ref{numcounts}). 

\section{Number counts}\label{numcounts}

\begin{figure*}
  \includegraphics[width=170mm]{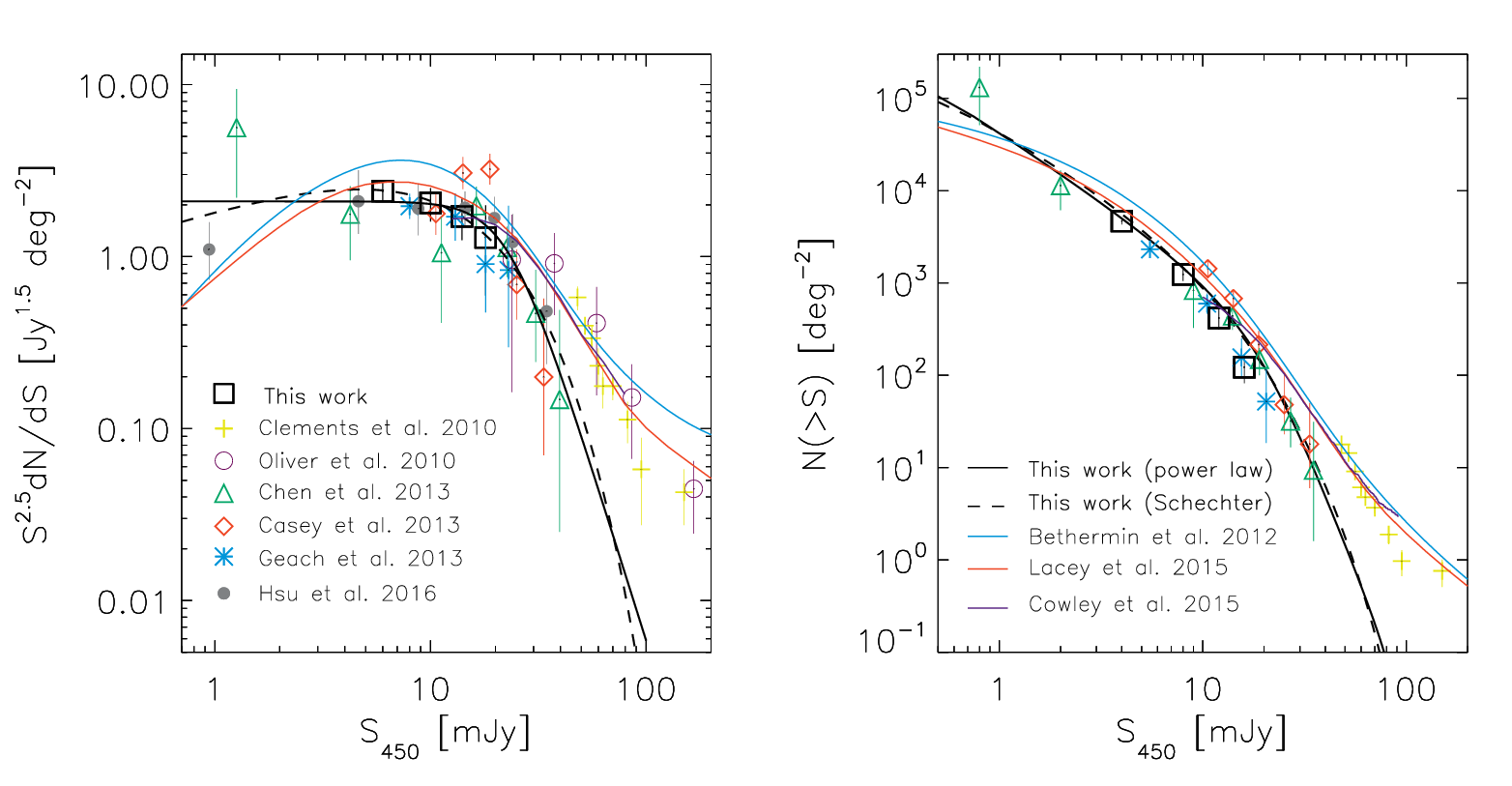}
  \includegraphics[width=170mm]{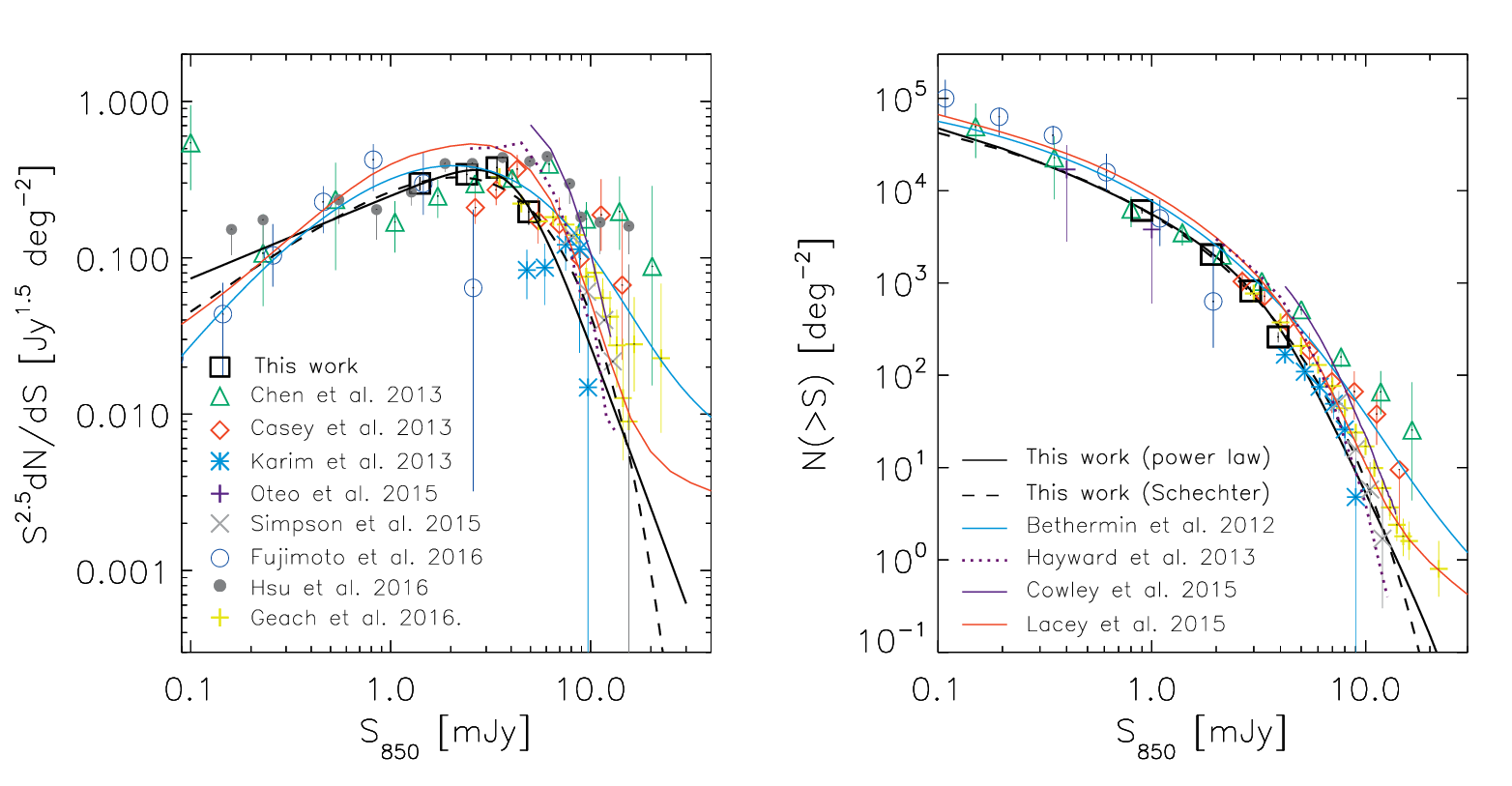}
  \caption{{\it Left:} Euclidean-normalized differential number counts of galaxies detected at 450 (top) and 850\! $\mu\rm{m}$ (bottom). 
  Error bars indicate the 68 per cent confidence intervals. The solid black line represents the best fit using a double--power law, 
  and the dashed black line is the best-fit using a Schechter-like function to the data described in \S\ref{numcounts}. 
  For comparison, we also plot the results from other studies and some theoretical predictions from the literature. 
  {\it Right}: Corresponding cumulative number counts at 450 (top) and 850\! $\mu\rm{m}$ (bottom) together with integrated fits.}
  \label{num_counts}
\end{figure*}

The cumulative number counts (also called `source counts') describe the number density of galaxies as a function of flux density. 
To derive this quantity, we adopt the standard bootstrap sampling method that has been extensively used  by different authors
(e.g. \citealt{2005MNRAS.357.1022C, 2006MNRAS.372.1621C}; \citealt{2009MNRAS.393.1573A, 2010MNRAS.401..160A}). 
While other techniques commonly used for the extraction of number counts can in principle estimate the counts at
fainter flux densities (for example the `{\it P(D)}' technique; \citealt{2009ApJ...707.1750P}; \citealt{2010MNRAS.409..109G}), it has been claimed that they are more 
dependent on the assumed model (see discussion in \citealt{2010MNRAS.405.2260S}),  require constant instrumental noise across the image 
(which is not our case) and a good understanding of the instrumental PSF and source clustering (e.g. \citealt{2014MNRAS.440.2791V}).
On the other hand, with the Bayesian approach, the estimated counts are only weakly dependent on the assumed model of the prior distribution 
(\citealt{2009MNRAS.393.1573A, 2010MNRAS.401..160A}). Since this method is described in detail in 
the aforementioned papers, we only briefly summarize it here.

Using the source catalogue constructed from all the source candidates with S/N\! $>3.0$ and following the procedure described in Section~\ref{sec:deboosting}, we
derive the PFD for each source candidate.
In each realization, we randomly assign flux densities to the sources in the catalogue according to their respective  PFD, 
and then the sources are binned to derive the differential ($dN/dS$) and cumulative [$N$($>S\rm$)] number counts, correcting each
bin by the corresponding completeness and dividing by the survey area.
To  avoid including a large number of false positives, we only include sources whose posterior flux distribution indicates a 
probability of less than 5 per cent of having a negative intrinsic flux, $P$($S<0$) $\le0.05$.
This process is repeated 500 times, also taking into account the error in 
completeness, in order to sufficiently sample the number count probability distribution. 

We calculate the number counts at 450\! $\mu\rm{m}$ for flux densities $\ge 4$\! mJy and at 850\! $\mu\rm{m}$ for flux densities $\ge 0.9$\! mJy. 
At lower flux densities, the level of completeness is too low ($\le 25$ per cent), and therefore, the errors could be large. These represent the
deepest number counts derived from single-dish telescope observations towards a blank-field (i.e. without the benefit 
of lens amplification). It is important to remark that the confusion noise in this map has been measured to be $\sigma_c\approx 0.4$\! mJy\! beam$^{-1}$ at 850\!
$\mu\rm{m}$, which is comparable to the instrumental noise in the map, and therefore should be taken into account.
Following the results of \citet{2010MNRAS.405.2260S} we have estimated the
completeness by inserting mock sources one at a time in the real flux maps (see \S\ref{sec_completeness}). This method has the advantage
of taking into account the confusion effects, since the sources are inserted in the real map, and does not over-predict the completeness, as 
when the sources are inserted into noise maps, as is commonly done in the literature.

Table~\ref{countstable} lists the resulting bin centres, differential and cumulative number counts, and 68 per cent confidence interval 
uncertainties. Figure~\ref{num_counts} shows the cumulative and differential number counts at each wavelength.

\begin{table*}
  \caption{Differential and cumulative number counts at 450 and 850\! $\mu\rm{m}$, calculated as described in Section~\ref{numcounts}, along with the best-fit
  parameters.}
  \begin{tabular}{ccccccccc}
    \hline
    \multicolumn{4}{c}{\underline{450\! $\mu\rm{m}$-number counts}}&&\multicolumn{4}{c}{\underline{850\! $\mu\rm{m}$-number counts}}\\
    {\it S} & d{\it N}/d{\it S} $^\dagger$  &{\it S}& {\it N} ($>${\it S})&& {\it S} & d{\it N}/d{\it S} $^\dagger$ & {\it S} & {\it N}($>${\it S})\\
    (mJy) & (mJy$^{-1}$deg$^{-2}$) & (mJy) & (deg$^{-2}$)&&(mJy) & (mJy$^{-1}$deg$^{-2}$) & (mJy) & (deg$^{-2}$)\\
    \hline
    6.0 & $873\pm 100$& 4.0 & $4742\pm 360$ &&1.4 & $4076\pm 400$& 0.9 & $6106\pm 450$ \\
    10  &$207\pm 45 $\!   & 8.0 & $1250\pm 186$ &&2.4 & $1210\pm 180$& 1.9 & $2030\pm 140$\\
    14  & $76\pm 20$ & 12 &  $423\pm 104$  &&3.4 & $ 559\pm 100 $ & 2.9 & $820\pm 90$ \\
    18  & $29\pm 16$ & 16 & $119\pm 40$  &&4.9 & $ 117\pm 20 $  & 3.9 & $261\pm 33$ \\
    \hline
    \multicolumn{9}{c}{\underline{Best-fit Schechter function using all data }}\\
    $N_0$ & $S_0$ & $\alpha$ & & & $N_0$ & $S_0$ & $\alpha$ &\\
    \hline
    $7100 \pm 600$ & $8.9 \pm 1.5$ & $2.6 \pm 0.2$ & & & $8300 \pm 300$ & $2.3 \pm 0.5$ & $2.6 \pm 0.8$ \\
      \multicolumn{9}{c}{\underline{Best-fit double-power law using all data}}\\
    $N_0$ & $S_0$ & $\alpha$ & $\beta$& & $N_0$ & $S_0$ & $\alpha$ & $\beta$\\
    \hline
    $600 \pm 140$ & $23 \pm 4$ & $2.5 \pm 0.2$ &$6.5 \pm 0.2$ & & $1900 \pm 600$ & $4.3 \pm 0.5$ & $2.0 \pm 0.4$ & $6.0 \pm 0.3$ \\
    \hline  
    \multicolumn{9}{c}{\underline{Best-fit Schechter function using only S2CLS data }}\\
    $N_0$ & $S_0$ & $\alpha$ & & & $N_0$ & $S_0$ & $\alpha$ &\\
    \hline
    $4400 \pm 2300$ & $11 \pm 4$ & $3.1 \pm 0.3$ & & & $16000 \pm 4000$ & $1.3 \pm 0.5$ & $1.7 \pm 0.8$ \\
    \hline  
    \multicolumn{8}{l}{$^\dagger$ The differential number counts reported in the table are not Euclidean-normalized. }
    \label{countstable}
  \end{tabular}
\end{table*}

We describe the differential number counts by a Schechter-like function of the form

\begin{equation}
  \frac{dN}{dS}=\left(\frac{N_0}{S_0}\right)\left(\frac{S}{S_0}\right)^{1-\alpha}\exp\left(-\frac{S}{S_0}\right).
\end{equation}

Alternatively, the number counts can also be fit with a double--power law described by
\begin{equation}
\frac{dN}{dS}=\frac{N_0}{S_0}\left[\left(\frac{S}{S_0}\right)^\alpha + \left(\frac{S}{S_0}\right)^\beta\right]^{-1}, 
\end{equation}
where $N_0$, $S_0$, $\alpha$ and $\beta$ describe the normalisation, break, and slope of the power laws, respectively. 

To determine the best-fit parameters we perfom a $\chi^2$ optimisation using a Levenberg--Marquardt algorithm.
In order to better determine the best-fit parameters, we include the results from other surveys in the fitting.
At 450\! $\mu\rm{m}$ we use all the SCUBA-2 measurements (\citealt{2013ApJ...776..131C}; 
\citealt{2013MNRAS.432...53G}; \citealt{2013MNRAS.436.1919C}), which cover a flux density range of $S_{450}\approx 1-40$\! mJy. 
We exclude results from {\it Herschel} surveys at higher flux densities, since these estimates are dominated by the lensed population.
At 850\! $\mu\rm{m}$ we complement our measurements at fainter flux densities with the results from SCUBA-2 observations 
in lensed fields (\citealt{2013ApJ...776..131C}). However, we exclude the estimates of \citet{2016ApJS..222....1F} because of the additional uncertainty in the 
1.2\! mm-to-850\! $\mu\rm{m}$ scale factor. At higher flux densities we complement our measurements with the results of ALMA observations 
(\citealt{2013MNRAS.432....2K}; \citealt{2015ApJ...807..128S}), and exclude the rest of the single-dish measurements, since effects of
blending are more significant at these high flux densities. 

The best-fit parameters for both models are listed in Table~\ref{countstable} and plotted in Figure~ \ref{num_counts}. The errors were estimated 
through Monte Carlo simulations. We find that both functional forms (Schechter-like and double--power law) produce similar fits. 

The last bin of our 450\! $\mu\rm{m}$ cumulative number counts lies below of our best-fit function. This is beacuse we did not find any source
with $S_{450}>18$\! mJy, most likely due to cosmic variance in our relatively small map.

\subsection{Comparison to other surveys}
We compare our number counts at both wavelengths to the results of previously published surveys in Figure~\ref{num_counts}.

At 450\! $\mu\rm{m}$ our results are in very 
good agreement with the counts from the previous S2CLS map in the COSMOS field \citep{2013MNRAS.432...53G}, which has similar depth and area. In the 
same field \citet{2013MNRAS.436.1919C} presented wider but shallower observations, which allowed them to estimate the number counts to 
brighter flux densities, and are also consistent with our values. 
\citet{2013ApJ...776..131C} and \citet{2016arXiv160500046H} combined data from cluster lensed 
fields and blank fields, measuring the counts over a wide flux density range. Our estimates are consistent with their results,
as well as with our extrapolation of the Schechter function. At flux densities above $\sim20$\! mJy, the number counts from our observations could be compared 
with the results from {\it Herschel} surveys, which mapped wider areas at 500\! $\mu\rm{m}$ to shallower depths. Our results are in 
agreement with the values reported by \citet{2010A&A...518L..21O} and \citet{2010A&A...518L...8C} at $\sim 20$\! mJy, where the distributions meet. 
At higher flux densities the counts estimated using {\it Herschel} are dominated by rare bright and lensed galaxies that our smaller area map cannot constrain.

At 850\! $\mu\rm{m}$ the measurements from our survey are in agreement with the values of \citet{2013ApJ...776..131C} and \citet{2016arXiv160500046H}, which came from
both lensed and blank fields, except for the brighter flux density bins ($\ga7$\! mJy), in which our extrapolation of the Schechter function lies
below their estimates. The same is true for the \citet{2013MNRAS.436.1919C} results. 
On the other hand, at such high flux densities our extrapolation of the Schechter 
function is in good agreement with recently results from follow-up ALMA observations of SMGs detected with single dish telescopes
(\citealt{2013MNRAS.432....2K}; \citealt{2015ApJ...807..128S}), and with
the number counts derived from $\sim5$ deg$^2$ SCUBA-2 observations (the largest and 
deepest single survey at 850\! $\mu\rm{m}$ so far; \citealt{2016arXiv160703904G}).
At fainter flux densities, our results overlap with the deep 870\! $\mu$m ALMA observations presented by \citet{2015arXiv150805099O} in excellent agreement, although
their uncertainties are large because of the small number of sources detected (11 sources in $\sim6$ arcmin$^2$ combining ALMA observations at different depths). 
At the same time, our estimations are consistent with the results at 1.2\! mm (scaled to 850\! $\mu\rm{m}$ by a factor of 2.3, assuming a typical SED at $z\sim2$) 
by \citet{2016ApJS..222....1F}, which include all the archival deep data at that time 
(including data from other deep ALMA observations, e.g. \citealt{2013ApJ...769L..27H}; \citealt{2014ApJ...795....5O}; \citealt{2015A&A...584A..78C}), compiling
a sample of 133 sources within $\sim10$ arcmin$^2$. 

Excluding the results from \citet{2013ApJ...776..131C} and \citet{2016arXiv160500046H}, which use the benefit of lensing amplification, this is the first time that
single-dish blank field observations have connected with the results from deep interferometric measurements. This results in a better understanding of 
the number counts towards fainter flux densities.

Since we have included previously reported number counts in our fits, may not be surprising that our results are in agreement with previous estimates of the 
number counts. To test if our data alone are in good agreement with previous results,
we again run the fitting procedure, but just using our measurements. Here we only apply the Schechter function since the double--power law has one more free
parameters, making it impractical to fit to just four bins. The best-fit parameters are listed in Table~\ref{countstable}. 
At 450\! $\mu\rm{m}$, although the parameters of the Schechter function are different, the fit is in very good agreement with the values in 
the literature, from about $1-40$\! mJy. At 850\! $\mu\rm{m}$, our measurements are in good agreement with the brighter flux density estimations, 
however, at fainter flux densities ($<0.5$\! mJy) our extrapolation of the Schechter function is below the measurements of \citet{2013ApJ...776..131C}
(although still consistent within the error bars), but in very good agreement with the ALMA
estimates of \citet{2015arXiv150805099O}.

\subsection{Comparison to models}
In this section, we compare the 450\,$\mu$m and 850\,$\mu$m number counts presented in this work to the results of recent 
galaxy formation models. \citet{2015arXiv150908473L}, presented a new version of the {\sc GALFORM} semi-analytical model, 
which includes improvements to the prescription for AGN feedback, disk--instability--driven starbursts, and stellar 
population models.  \citet{2015MNRAS.446.1784C} implemented also the effect of the beam--size on the observed number 
counts on the GALFORM model to study the possible bias introduced by the source blending of individual sources. However,  
they conclude that the beam size at 450\! $\mu\rm{m}$ does not produce any significant enhancement of the source density.
In Figure \ref{num_counts}, it can be seen that the number counts predicted by GALFORM appear to be in broad agreement with the 
results presented here within the  450\,$\mu$m flux density range of $S_{450}$\,=\,3--30\,mJy. 
However, while the shape of the number counts are broadly similar we identify a small offset between the observed counts and
the theoretical predictions, where the GALFORM source density lie a factor of $\sim1.3$ above of our integrated measurements. In contrast, at the faintest flux
the model underpredict the number of sources (a factor $\sim2$ in the cumulative number counts at $S_{450}\approx0.6$\! mJy). At higher flux densities, the number 
counts become dominated by the lensed population that, as described in the previous section, our survey has insufficient 
area coverage to constrain accurately. We also compare to the \citet{2012ApJ...757L..23B} model, which is based on the evolution of the main-sequence of star-forming 
galaxies and a second population of starburst galaxies, assuming some spectral energy distribution templates. The results of 
\citet{2012ApJ...757L..23B} are similar to the {\sc GALFORM} model and therefore are in reasonable agreement with our measurements, 
albeit with a marginally larger offset of $\sim1.7$ above our cumulative counts (Figure~\ref{num_counts}.)

At 850\! $\mu\rm{m}$ the number counts predicted by the GALFORM model (\citealt{2015arXiv150908473L}) also follow the behavior of our best-fit source
counts at flux density of $S_{850}$\,=\,0.2--10\,mJy. Although, we again note that the theoretical counts are 
systematically a factor $\sim1.5$ above the observed values. 
Interferometric observations have shown that the source blending at this wavelength is more important due to the  larger beam--size 
(e.g. \citealt{2011ApJ...726L..18W}; \citealt{2013ApJ...768...91H}). This effect is also predicted by  \citet{2015MNRAS.446.1784C} 
when taking into account the coarser angular resolution in the GALFORM model. The \citet{2012ApJ...757L..23B} predictions are also in agreement 
(but also a factor $\sim1.4$ above our 
best-fit integrated functions), nevertheless, when compared with our estimates and the \citet{2016arXiv160703904G} results, the
model overpredicts the counts at $S_{850}\ga 8$\! mJy, for example, by a factor of $\sim5$ in the cumulative counts at $S_{850}\approx16$\! mJy 
(see Figure~\ref{num_counts}). Finally, we also compare our measurements to the model of \citet{2013MNRAS.434.2572H} based 
on the {\it bolshoi} cosmological simulation, which also takes into account the blending in single-dish observations. The 
predictions for a 15$''$ beam are  consistent with our number counts, however, this model predicts that the 
multiplicity caused by blending increases the number counts by more than an order of magnitude. This has been ruled out by
recent interferometric results which found that the number counts are boosted by only 20 per cent at $S_{870}>7.5$\! mJy, 
and 60 per cent at $S_{870}>12$\! mJy (\citealt{2015ApJ...807..128S}).

\section{Contribution to the Cosmic Infrared Background}
\subsection{The resolved CIB}\label{sec:CIB}
Once we have extracted the point sources with S/N\! $>3.5$ from our maps, we can estimate the contribution of  
these sources to the CIB at each wavelength, corrected for completeness, by integrating the number counts 
above our flux density limit. To do this we integrate the best-fit number counts (those derived when used all the data) at $S_{450}>4.0$\! mJy and 
$S_{850}>0.9$\! mJy. At these flux densities the integration of both the double-power law and the Schechter
function  give us the same results. The integrated intensities of our detected galaxies are 
$I_\nu(450\mu\rm{m})=(0.13\pm0.03$\! MJy\! sr$^{-1}$ and   $I_\nu(850\mu\rm{m})=0.04\pm0.01$\! MJy\! sr$^{-1}$, 
which correspond to $28\pm13$  and $28\pm14$ per cent, respectively, of the CIB 
measured by the COBE Far-Infrared Absolute Spectrophotometer (FIRAS, \citealt{1998ApJ...508..123F}), where the uncertainties are dominated by those 
of the total CIB values. Extrapolating the number counts below our detection limits, we find that the total CIB is resolved at $S_{450}\approx0.6$\! 
mJy and $S_{850}\approx0.02$\! mJy, respectively.

There have been other measurements of the total values of the CIB based on both direct measurements, for example of COBE 
Difuse Infrared Background Experiment (DIRBE, \citealt{1998ApJ...508...25H}) and COBE FIRAS (\citealt{1999A&A...344..322L}), and integrated galaxy-counts
derived from {\it Spitzer} and {\it Herschel} (e.g. \citealt{2006A&A...451..417D}; \citealt{2011A&A...532A..49B}; \citealt{2012A&A...542A..58B}; see also
recent compilation by \citealt{2016arXiv160501523D}). However,
the uncertainties remain large, and therefore, better estimates of the absolutes CIB values are still required to better constrain
the fraction contributed by these galaxies. Here we adopt the values of \citet{1998ApJ...508..123F} in order to consistently compare our results to previous 
similar studies (e.g. \citealt{2015ApJ...809L..22V}).

To measure the contribution to the CIB by galaxies fainter than our detection threshold, 
we stack the maps at the positions of 24\! $\mu\rm{m}$ galaxies selected from a {\it Spitzer} catalogue
in the EGS field \citep{2011ApJS..193...13B, 2011ApJS..193...30B}.  This technique has been implemented before by different
authors (e.g. \citealt{2009ApJ...707.1729M};  \citealt{2012A&A...542A..58B}; \citealt{2013ApJ...779...32V}; and references therein), and used
by \citet{2013MNRAS.432...53G} in the COSMOS/S2CLS survey. First, we  
remove point sources detected with S/N$>3.5$ from the maps using a point spread function (PSF) 
normalized to the flux of each individual source in our catalogue. Then, we subtract the mean of the SCUBA-2 maps, which yields a 
residual map, where the flux corresponds to sources that are not present in our catalogues (in addition to noise).
Finally, the 450 and 850\! $\mu\rm{m}$ maps are stacked at the position of the 24\! $\mu\rm{m}$ sources, averaging the flux at 
each position. 

\begin{figure}
  \includegraphics[width=\columnwidth]{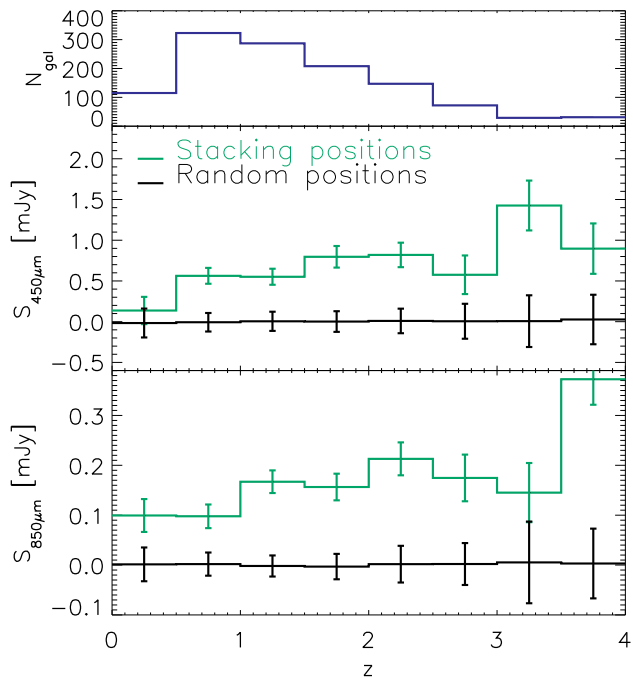}
  \caption{Stacked average flux density at 450 and 850\! $\mu\rm{m}$ (green line in the {\it middle} and {\it bottom} panels, respectively) as a function of
  redshift at the position of the 24\! $\mu\rm{m}$-selected galaxy sample. The top panel represents the number of galaxies in bins of redshift used in the stacking procedure.
  The black lines represent the results of null tests, in which the stacking was done at random positions, finding an average stacked flux of zero. }
  \label{stacked_sample}
\end{figure}

Since the 24\! $\mu\rm{m}$ source catalogue includes photometric redshifts, we can estimate the stacked flux density as a function 
of redshift, as shown in Figure~\ref{stacked_sample}. The uncertainty is estimated as $\sigma/\sqrt{N}$, with $\sigma$
is the standard deviation in the stack and $N$ the sample size (which is represented in the top panel of Figure~\ref{stacked_sample}). To ensure that the
recovered stacked flux comes from the 24\! $\mu\rm{m}$ sample and not from noise or other 
contaminants, we repeat the same procedure but at random positions, conserving the number of
stacked positions in each redshift bin. As shown in Figure~\ref{stacked_sample}, the average stacked flux from the random positions is zero, which means
that the recovered fluxes are actually associated with the 24\! $\mu\rm{m}$ population.

The total intensities recovered by the stacking are $I_\nu(450\mu\rm{m})=0.13\pm0.01$ MJy sr$^{-1}$ and $I_\nu(850\mu\rm{m})=0.03\pm0.003$ MJy sr$^{-1}$,
which corresponds to $28\pm11$ and $21\pm9$ per cent of the CIB, respectively, including the uncertainties 
in the absolute CIB values

The contribution of the directly detected sources and the stacking of 24\! $\mu\rm{m}$ galaxies amount a total 
$0.26\pm0.03$ and $0.07\pm0.01$  MJy sr$^{-1}$ at 450 and 850\! $\mu\rm{m}$, respectively, which correspond to $\sim60\pm20$ and $50\pm20$ per cent of the total CIB measurements.
This is in excellent agreement with the estimations by \citet{2012A&A...542A..58B}, who using the 24\! $\mu\rm{m}$ sources as priors, estimating 
that the emission of galaxies down to $\sim2$\! mJy at 500\! $\mu\rm{m}$ contributed  $\sim 55$ per cent of the CIB.

\begin{figure*}
  \includegraphics[width=90mm]{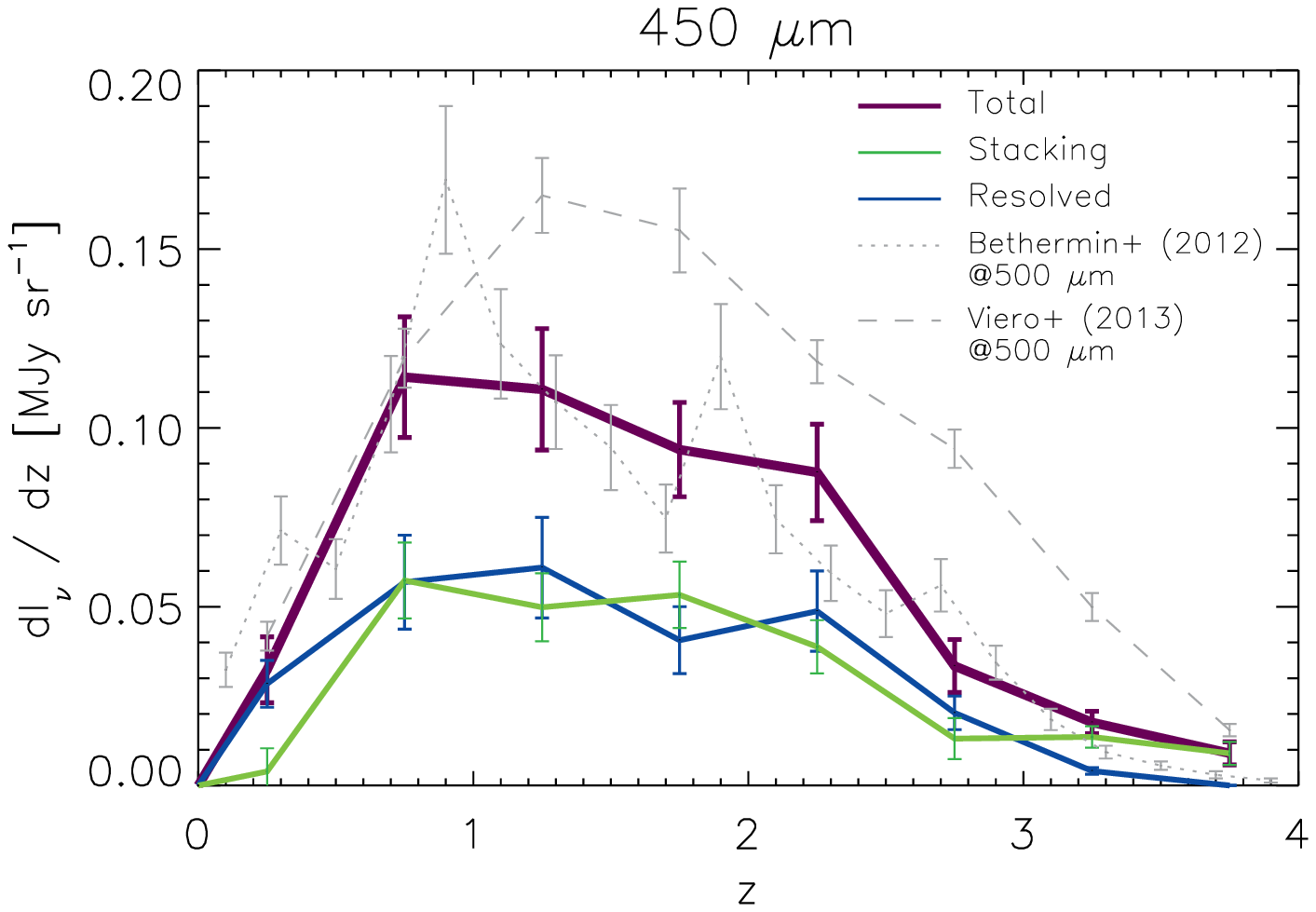}\includegraphics[width=90mm]{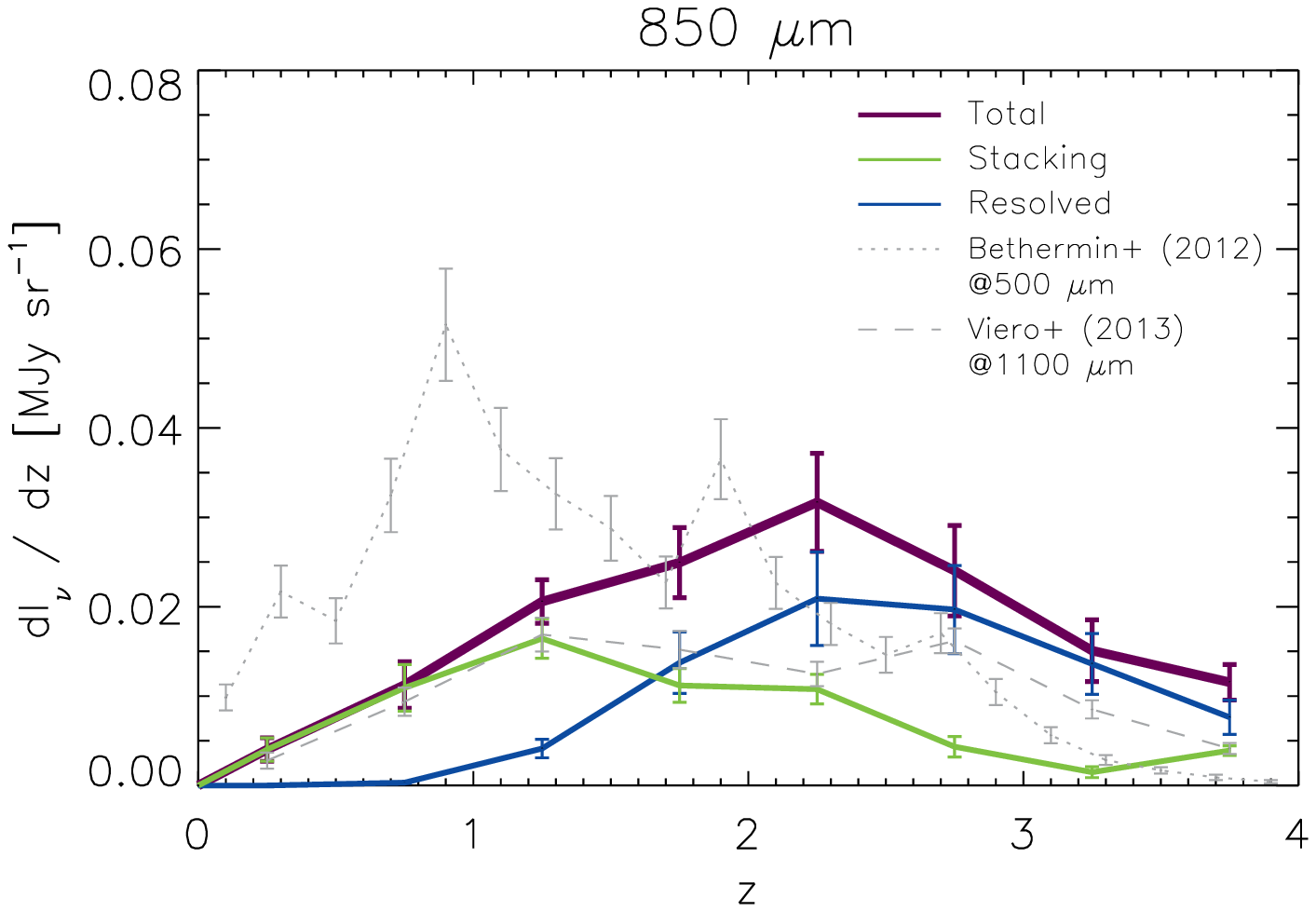}
  \caption{Redshift distributions of the recovered intensity, $dI_\nu/dz$, at 450\! $\mu\rm{m}$ (left) and 850\! $\mu\rm{m}$ (right). 
  The blue line represents the contribution of the point sources directly detected in the SCUBA-2 maps and the green line 
  represents the contribution from stacking 24\! $\mu\rm{m}$-detected galaxies. The total contribution is plotted as a purple line. 
  For comparison we plot the results from \citet{2012A&A...542A..58B} at 500\! $\mu\rm{m}$ and from \citet{2013ApJ...779...32V}  at  500 and 1100\! $\mu\rm{m}$. 
  The 500\! $\mu\rm{m}$ values have been multiplied by a factor of 1.2 and 0.36 to scale their total intensities to match our 450 and
  850\! $\mu\rm{m}$ maps, while the  1100\! $\mu\rm{m}$ are scaled by 1.75 to match our 850\! $\mu\rm{m}$ measurements, according to the absolute values measured by \citet{1998ApJ...508..123F}, 
  }
  \label{CIB_450}
\end{figure*}

An important concern in the stacking technique is the possible bias due to clustering, which can result in a boosted average flux
density arising from faint, companion (or clustered) galaxies (e.g. \citealt{2008MNRAS.386.1907S}; \citealt{2010AJ....139.1592K};
\citealt{2013MNRAS.429.1113H}). This effect increases with
the size of the beam. However, using a 24\! $\mu\rm{m}$ catalogue and the {\it Herschel} maps, \citet{2012A&A...542A..58B} found
that this bias is only $5-7$ per cent at 250\! $\mu\rm{m}$, which has ${\rm FWHM}=18.1$ arcsec. Our 850\! $\mu\rm{m}$ beam 
has ${\rm FWHM}=14.5$ arcsec, and therefore,  the expected bias due to clustering should be less than
5 per cent (and even less at 450\! $\mu\rm{m}$ with a beamsize of ${\rm FWHM}=8$ arcsec). 
Given the uncertainties in our measurements and the absolute values of the CIB, we have not included any clustering correction in our estimations

Considering that roughly half of the CIB is still missing at both wavelengths it is important to discuss the possible origin for this missing fraction. 
The first point to consider is that we do not correct our stacking measurements for the completeness of the 24\! $\mu\rm{m}$-selected catalogue. As 
discussed in earlier studies (e.g. \citealt{2012A&A...542A..58B}), the stacking of an incomplete catalogue  biases the result, missing a significant fraction of 
the intensity. \citet{2013A&A...557A..66B} show that an $S_{24\mu\rm m}>80$\! $\mu\rm{Jy}$ selection could miss up to half of the intensity expected from all galaxies, 
which could account for the remaining CIB.

Actually, \citet{2015ApJ...809L..22V} studied the contribution of galaxies which are not detected in current 
near-infrared surveys, for example, very low-mass or very dusty galaxies, but which are correlated (i.e. clustered) with the detected galaxies.  
To account for these uncatalogued sources they intentionally smoothed and stacked {\it Herschel} observations. 
They found that the contribution of these galaxies is very important, especially 
in the $1<z<4$ range (where the completeness is low), and could fully explain the rest of CIB at 250 -- 500\! $\mu\rm{m}$. 
However, at longer wavelengths (i.e. $\lambda\ga850$\! $\mu\rm{m}$), deep interferometric observations  (e.g \citealt{2014ApJ...789...12C}; 
\citealt{2016ApJS..222....1F}) have shown that only $\sim50$ per cent of the faint submillimetre galaxies ($S_{850}\la2$\! mJy) are detected in deep 
optical/NIR surveys, suggesting that many of these sources, which also contribute to the CIB, are high redshift galaxies ($z\ga3$).

\subsection{The redshift distribution of the recovered CIB}

As described above, the stacking analysis was performed in bins of redshifts, which allowed us to estimate the contribution 
of these galaxies as a function of redshift. The green line in Figure~\ref{CIB_450} represents the redshift distribution 
of the recovered intensity from the stacking of 24\! $\mu\rm{m}$ prior positions at 450 and 850\! $\mu\rm{m}$. 

To estimate the redshift distribution of the intensity produced by the 
galaxies formally detected in our SCUBA-2 maps we adopt previously published redshift distributions of
similar populations. For the 450\! $\mu\rm{m}$-detected galaxies we use the distribution of
\citet{2013MNRAS.436..430R}, which comprises photometric redshifts of 450\! $\mu\rm{m}$-selected galaxies detected 
in deep SCUBA-2 observations ($\sigma_{450}=1.5$\! mJy), similar to the depth of our map, and therefore, galaxies 
with similar flux densities. This redshift distribution shows a broad peak in the range $1<z<3$, and a median of $z=1.4$ 
\citep{2013MNRAS.436..430R}. For the 850\! $\mu\rm{m}$ galaxies we consider the redshift distribution of \citet{2016ApJ...820...82C},
derived from a large sample of SCUBA-2-detected galaxies with $S_{850}>1$\! mJy, which has a median
redshift of $2.6 \pm 0.1$, in consistency with the deep ($\sigma_{850}=0.25$\! mJy) S2CLS COSMOS map (\citealt{2016MNRAS.458.4321K})
and with previous observations of brighter sources (e.g. \citealt{2005ApJ...622..772C}). 
Figure~\ref{CIB_450} indicates the intensity as a function
of redshift which is contributed by the galaxies detected in our maps at both wavelengths, this being the result of scaling the redshift 
distribution to the corresponding intensity. The redshift distribution of the total intensity (stacked plus directly detected sources) 
is also and is reported in Table~\ref{diff_CIB}.

As shown in Figure~\ref{CIB_450} the redshift distribution of the recovered CIB is different at the two wavelengths, 
with sources at higher redshifts contributing more at 850\! $\mu\rm{m}$. 
At 450\! $\mu\rm{m}$ our results show a peak at $z\sim1$,
in very good agreement with the measurements at 500\! $\mu\rm{m}$ by \citet{2012A&A...542A..58B}, which come 
from 24\! $\mu\rm{m}$ catalogues, and with \citet{2013ApJ...779...32V}, who stacked {\it K} band--selected sources in {\it Herschel} maps.  
Their values have been plotted in Figure~\ref{CIB_450} (left panel)
multiplied by a factor of 1.2 to scale the 500\! $\mu\rm{m}$ estiamtes to our 450\! $\mu\rm{m}$ measurements, 
according to the flux density ratio of the CIB spectrum (\citealt{1998ApJ...508..123F}). At 850\! $\mu\rm{m}$ the 
redshift distribution of the recovered intensity peaks at $z\sim2$ (see right panel of Figure~\ref{CIB_450}). 
We have again plotted the values of \citet{2012A&A...542A..58B} (scaled to 850\! $\mu\rm{m}$ by a factor of 0.36) as
a comparison. As shown in this Figure, the \citeauthor{2012A&A...542A..58B} redshift distribution is inconsistent with our values, clearly 
showing that the contribution of higher redshift sources is more important at longer wavelengths. Our estimated redshift distribution is, on the other hand, 
similar to the one found by \citet{2013ApJ...779...32V} but at 1100\! $\mu\rm{m}$. Their results (scaled at 
850\! $\mu\rm{m}$ by a factor of 1.75) are also plotted in Figure \ref{CIB_450}. 

Our results, highlight that the redshift distribution of the CIB depends on wavelength, with the peak shifting to higher redshifts for longer
wavelength bands, and vice versa, as also measured by \citet{2009ApJ...707.1729M}, \citet{2011A&A...532A..49B}, \citet{2012A&A...542A..58B}, and 
\citet{2013ApJ...779...32V}. This is also consistent with the predictions of different theoretical models of the CIB
(e.g. \citealt{2009ApJ...701.1814V}; \citealt{2013A&A...557A..66B}; \citealt{2013ApJ...772...77V}; \citealt{2016arXiv160402507M}). This  selection
effect has also been observed in the redshift distributions of submillimetre galaxies selected at different wavelengths, in which shorter wavelengths select lower-$z$ galaxies 
(and hotter sources), and can be explained with a single population of galaxies (\citealt{2014MNRAS.443.2384Z}, \citealt{2015A&A...576L...9B})

On the other hand, \citet{2015MNRAS.446.2696S}, using {\it Planck} data, have measured a peak at $z\sim1.2$ for 350, 550 and 850\! $\mu\rm{m}$ 
observations, with no obvious wavelength dependence, however, the uncertainties around the peak are large enough to
hide or mask out the possible evolution as a function of wavelength, as discussed by these authors.

As  discussed above, most of the remaining CIB is expected to be emitted by obscured or low-mass galaxies at $z<4$ that are not present in the  
optical/NIR catalogues due to incompleteness, as suggested by the recent work of \citet{2015ApJ...809L..22V}.
However, they also found tentative evidence of
higher redshift ($z>4$) contributions to the CIB at 500\! $\mu\rm m$. This population of high-$z$ galaxies, which is unreachable with the current  
optical/NIR surveys (e.g. \citealt{2016arXiv160100195K}), may also contribute to the CIB (although to a lesser extent), especially at longer wavelengths.
Actually, \citet{2016ApJS..222....1F} found that the full CIB at 1.2\! mm is explained by sources with $S_{1.2}>0.02$ mJy, but only half of the 
faintest galaxies ($0.02<S_{1.2}<1.2$\! mJy) has an optical/NIR counterpart, and none has a radio, suggesting a high-redshift ($z>3$) nature.

\begin{table}
  \centering
  \caption{Differential recovered intensity as a function of redshift at 450 and 850\! $\mu\rm{m}$ (as plotted in Figure~\ref{CIB_450}).}
    \begin{tabular}{ccc}
    \hline
    Redshift & 450\! $\mu\rm{m}$ & 850\! $\mu\rm{m}$\\ 
	    &  \multicolumn{2}{c}{($\times 10^{-2}$ MJy sr$^{-1}$)}\\
    \hline
    $0.0 < z < 0.5$ & $3.0^{+1.3}_{-1.1}$ & $0.3^{+0.1}_{-0.1}$ \\
    $0.5 < z < 1.0$ & $12.3^{+2.5}_{-2.0}$ & $0.9^{+0.3}_{-0.3}$ \\
    $1.0 < z < 1.5$ & $12.3^{+2.5}_{-2.0}$ & $2.3^{+1.1}_{-0.6}$ \\
    $1.5 < z < 2.0$ & $9.0^{+1.8}_{-1.5}$ & $3.0^{+2.6}_{-1.3}$ \\
    $2.0 < z < 2.5$ & $9.4^{+2.1}_{-1.6}$ & $3.5^{+3.0}_{-1.4}$ \\
    $2.5 < z < 3.0$ & $4.4^{+1.1}_{-0.9}$ & $2.2^{+2.2}_{-1.1}$ \\
    $3.0 < z < 3.5$ & $1.8^{+0.4}_{-0.4}$ & $1.1^{+1.3}_{-0.6}$ \\
    $3.5 < z < 4.0$ & $1.0^{+0.3}_{-0.3}$ & $0.7^{+0.6}_{-0.3}$ \\
    \hline
    \label{diff_CIB}
    \end{tabular}
\end{table}

\section{Summary}

We have presented deep SCUBA-2 observations at 450 and 850\! $\mu\rm{m}$ in the EGS field as part of the SCUBA-2 Cosmology Legacy Survey. This survey, together 
with other similarly deep S2CLS maps, represents one of the deepest blank-field observations achieved with single-dish telescopes, reaching a depth of  $\sigma_{450}=1.2$\! mJy 
and $\sigma_{850}=0.2$\! mJy, respectively. Using 57 sources detected above $3.5\sigma$ at 450\! $\mu\rm{m}$ and 90 at 850\! $\mu\rm{m}$ we estimate the 
number counts and the contribution of these galaxies to the cosmic infrared background. 

Our number counts at 450\! $\mu\rm{m}$, at a flux density limit of $S_{450}>4.0$\! mJy, are in good agreement with the previous estimation by \citet{2013MNRAS.432...53G}
derived from S2CLS observations in the COSMOS field. Our result is also consistent with previous shallower observations in blank and gravitational lensed fields.

At 850\! $\mu\rm{m}$ our results are the first number counts reported from the deep tier of the S2CLS, and therefore these represent the deepest number counts derived
from single-dish blank-field observations using only directly-detected sources. Our estimations are in agreement with the number counts achieved through the benefit
of gravitational lensing and with the recent results from interferometric observations with ALMA. 

We have also estimated the contribution of the detected sources to the CIB and the contribution
of 24\! $\mu\rm{m}$-detected galaxies throughout a stacking technique, which give a total of  
$0.26\pm0.03$ and $0.07\pm0.01$ MJy sr$^{-1}$, at 450\! $\mu\rm m$ 
and 850\! $\mu\rm m$, respectively, corresponding to $60\pm20$ and $50\pm20$ per cent 
of the CIB. Using the photometric redshifts available for the 24\! $\mu\rm{m}$-detected sample
and previously published redshift distributions of the 450 and 850\! $\mu\rm{m}$ blank-field 
population, we decompose this emission into bins of redshift, finding an evolution of the redshift 
distribution of the recovered CIB as a function of wavelength, which peaks at $z\sim1$ at  450\! $\mu\rm{m}$,
whereas at 850\! $\mu\rm{m}$ it peaks at $z\sim2$, in agreement 
with theoretical models and previous observations.

The remaining CIB is expected to be emitted by galaxies that are too faint at 24\! $\mu\rm{m}$ to have been detected
in {\it Spitzer} surveys, as discussed by other authors, although a contribution of high-redshift ($z>4$) 
galaxies could also be important, especially at the longer wavelength.

\section*{Acknowledgements}

{\small
We would like to thank the referee, Steve Eales, for a helpful report which has improved the clarity of the paper.

This research has been supported by Mexican CONACyT research grant
CB-2011-01-167291. JAZ is also supported by a CONACyT studentship. RJI acknowledges support from ERC 
in the form of the Advanced Investigator Programme, 321302, COSMICISM.
The James Clerk Maxwell Telescope has historically been operated by the Joint Astronomy Centre on behalf of the Science
and Technology Facilities Council of the United Kingdom, the National Research Council of Canada, and
the Netherlands Organisation for Science Research. Additional funds for the 
construction of SCUBA-2 were provided by the Canada Foundation for Innovation. This work is based in part
on observations made with the {\it Spitzer Space Telescope}, which is operated by the Jet Propulsion Laboratory, 
California Institute of Technology under a contract with NASA.}




\bibliographystyle{mnras}
\bibliography{biblio} 




\appendix

\section{Source catalogues}
The catalogues will be released in the published version.
\bsp	
\label{lastpage}
\end{document}